# The AI Security Zugzwang

Lampis Alevizos
*Volvo Group*
*Amsterdam, The Netherlands*
lampis@redisni.org

*Abstract* — In chess, zugzwang describes a scenario where any move worsens the player's position. Organizations face a similar dilemma right now at the intersection of artificial intelligence (AI) and cybersecurity. AI adoption creates an inevitable paradox: delaying it poses strategic risks, rushing it introduces poorly understood vulnerabilities, and even incremental adoption leads to cascading complexities. In this work we formalize this challenge as the AI Security Zugzwang, a phenomenon where security leaders must make decisions under conditions of inevitable risk. Grounded in game theory, security economics, and organizational decision theory, we characterize AI security zugzwang through three key properties, the forced movement, predictable vulnerability creation, and temporal pressure. Additionally, we develop a taxonomy to categorize forced-move scenarios across AI adoption, implementation, operational and governance contexts and provide corresponding strategic mitigations. Our framework is supported by a practical decision flowchart, demonstrated through a real-world example of Copilot adoption, thus, showing how security leaders can manage zugzwang positions balancing risk and innovation.

*Index Terms* — **AI cybersecurity, zugzwang, security decision-making, cybersecurity strategy, forced security moves, innovation.**

## I. INTRODUCTION

In late 2023, organizations worldwide found themselves in a dangerous position regarding AI adoption. The release of GPT-4 and similar large language models (LLMs) created enormous pressure to integrate AI capabilities into business operations. A survey of 53 Chief Information Security Officers (CISOs) [1] across US, Europe and Australia, reported they feel forced to adopt AI for competitive reasons, or to simply cope up with business requests by executive management, yet each adoption decision introduces new attack surfaces that nor themselves neither their teams fully understand as of today. Thereby, the integration of AI in enterprise environments has reached a critical inflection point. AI adoption started as a strategic choice, now it appears as an operational imperative, driven by competitive pressure, business operational efficiency demands or cybersecurity efficiency improvements [2]. A recent research report by MIT & BCG [3] showed that almost 85% of enterprises now view AI adoption as essential for maintaining competitive advantage, however, this adoption comes with unprecedented security challenges. Several scholars [4] [5] [6] as well as industry experts [7] [8] [9] have published research that shows AI systems introduce unique security vulnerabilities that go beyond standard threat models such as model poisoning attacks, extraction of training data through inference attacks and others. Thus, the technical security implications of this forced AI adoption are important because organizations nowadays must (1) defend AI systems, (2) defend from AI systems, (3) defend with AI systems, but whose behaviour, specifics and impacts they may not fully understand or are able to verify. Additionally, research [10] [11] demonstrates that AI models can be compromised through sophisticated adversarial attacks, and even seemingly secure implementations can leak sensitive information through model memorization [12]. Therefore, these existing technical challenges are amplified by organizational issues, namely, security teams must protect AI despite lacking established frameworks, standards, or even sufficient expertise in AI security [13].

This situation has created what we term "AI Cybersecurity Zugzwang", which means, positions where security leaders are forced to make security moves regarding AI, which will inevitably create new vulnerabilities or expose them to new risks. The term "zugzwang" however, is not new, rather it is borrowed from chess theory and widely used in chess to describe scenarios where every possible move worsens a player's position. In our case and considering the context of AI security, such positions arise when organizations must make decisions about AI adoption, implementation, or operation, knowing that each choice introduces new risks, while accepting or delaying the decisions creates different but equally significant risks.

Although extensive research exists on organizational technology adoption [14] [15], as well as industry [16] and government entities actively researching AI security [17], there is still a critical gap in understanding the forced-move situations that organizations face at the intersection of AI and cybersecurity. The existing literature primarily focuses on either technical security aspects of AI systems [18] [19] discussing technical vulnerabilities and defences, or organizational decision-making in AI technology adoption [20] [21], which tends to assume organizations have the luxury of strategic choice in their adoption decisions. However, the remaining research gap appears when addressing the security dilemmas created when organizations are required to make AI-security-related decisions, under the pressure of time.

This research gap is significant, considering the current state of AI adoption in enterprise environments, and the situations that CISOs and security teams are dealing with. Traditional security frameworks and risk assessment models are still applicable to a certain degree, given that AI can be viewed or treated as software, but they are proving inadequate [22] when organizations are evaluating trade-offs between immediate AI adoption risks and the, potentially, greater risks of delayed adoption. Another recent study on AI governance [23], highlights that organizations lack theoretical frameworks for understanding these forced-move scenarios, which in turn leads to suboptimal security related decisions. Therefore, the absence of formal models to analyse such situations leaves security leaders without structured approaches to identify, evaluate, and manage these challenges. Thus, we form the following research question:

*RQ1: How do organizations enter AI security zugzwang positions, what characterizes these positions, and how can security leaders effectively manage them?*

Addressing **RQ1**, we make the following contributions to both the theoretical understanding and practical management at the intersection of AI and cybersecurity:

**(1)** We introduce and characterize the concept of AI Security Zugzwang, thus, a novel theoretical framework for cybersecurity leaders to understand forced-move scenarios in cybersecurity. Our model considers situations where organizations must make security moves despite knowing these moves will create new vulnerabilities, as opposed to existing frameworks that treat security decisions as strategic choices [24].

**(2)** We develop a taxonomy of AI security zugzwang positions and organizational decision tactics. The taxonomy identifies distinct categories of forced-move scenarios, the causing conditions and their security implications. This contribution extends existing work on security decision-making [25], as it specifically addresses the limitations and adoption pressures introduced by AI.

**(3)** We provide practical strategic mitigations and position management, alongside a decision flowchart to help security leaders identify and enhance their decision making under zugzwang positions, based in game theory, the theoretical foundation section and empirical analysis. This addresses the gap in the existing traditional risk assessment methods that do not consider [22] the nuanced and oftentimes paradoxical nature of AI security decisions.

The remainder of this paper is organized as follows: in Section II we discuss the research methodology, next, in Section III we discuss the existing literature based on the combination of game theory, security economics, and organizational decision theory. We introduce our formal characterization and framework of AI security zugzwang and the methodology to identify and analyse such positions. Next, Section IV presents the taxonomy of zugzwang positions, supported by characteristics and contributing factors. Section V discusses the strategic mitigations, position management and details the decision flowchart using a real-world case of Copilot. Section VI examines implications for security practice and policy, while Section VII proposes future research directions. Lastly, in Section VIII we draw the conclusions of this work.

## II. RESEARCH METHODOLOGY

We followed the principles of theory building [26] to develop the AI Security Zugzwang framework and taxonomy within three phases. Phase 1 consists of mapping Z x S to C. Z represents zugzwang characteristics, S security decision scenarios, and C the identified commonalities. This first phase is the conceptual development where we analysed zugzwang positions in chess to identify their defining characteristics, with special focus on the forced-move nature and inevitable position deterioration that characterizes these situations. Parallel analysis of security decision-making scenarios in AI adoption and implementation showed several similarities, specifically regarding the forced nature of decisions and the absence of purely beneficial moves, which resonates with existing empirical evidence.

Phase 2 is the theoretical synthesis. It is captured as C → {T1, T2, T3}, where C is decomposed across the three theoretical domains T1, T2, T3, namely, game theory, security economics and organizational decision theory respectively. We started with game theory to understand the defender's choice limitations, and how standard assumptions about position preservation fail in AI security scenarios. Then, we analysed the security economics specifically on forced-choice scenarios and trade-offs, where the research gap was highlighted when security leaders address the inevitable security deterioration. Lastly, organizational decision theory showed why conventional security decision making is inadequate in such scenarios, under the pressure of time.

The final phase 3 is captured as {T1, T2, T3} → F, where F represents the resulting frameworks properties and taxonomy. The framework properties emerged from recurring patterns in security decision constraints appearing in the literature, and taxonomy development is the result of the systematic categorization of decision scenarios. The validation was done through cross referencing against documented AI security challenges and application to real-world implementation cases.

## III. THEORETICAL FOUNDATION

### A. Game Theory and Zugzwang

Extensive literature exists [27] [28] [29] on game theoretical approaches to cybersecurity , nonetheless, the application of game theory in cybersecurity is usually focused on attacker-defender interactions [28]. Our work, however, extends this to organizational decision-making under security constraints or under the pressure of time such as adoption timelines. Much like in chess, where players are forced to make decisions for the security of the pieces under the pressure of time. In classical security games (e.g. Stackelberg security games), defenders can choose not to move, or simply the "do nothing for now" option allows defenders to weigh the risks of different actions or conserve resources [30]. On the contrary, the adoption of AI systems and the threat events that they face, make inaction less viable, thus, requiring timely decisions to maintain the competitive advantage despite potential risks.

The concept of zugzwang and its application to cybersecurity decision-making is a novel theoretical extension because the traditional game theory in cybersecurity assumes that defenders can maintain current security positions [31], but in the context of AI adoption and security, maintaining the status quo means our position is deteriorating. We found this notion to be in alignment with recent research on dynamic security games [32], where the defender's position naturally degrades over time without active intervention. For instance, in AI model deployment, the initial position of a security team is to have fully validated model outputs, but this position degrades as the model encounters new types of inputs and edge cases. Therefore, without continuous monitoring and revalidation, the team cannot maintain confidence in the model's security properties. This is similar to how a chess player's initially secure position deteriorates when forced to make suboptimal moves under the pressure of time. Moreover, the degradation accelerates when adversaries probe the model with sophisticated attacks while the security team remains in their initial defensive posture.

### B. Security Economics and Forced Moves

This is another domain with well-established literature [33] [34]. Cost-benefit analysis is a fundamental and rational principle within cybersecurity economics [35]. However, in AI security zugzwang positions, new scenarios are introduced where all available choices introduce significant negative implications, thus, a new challenge. For instance, when organizations must decide on implementing AI-powered code analysis tools, they face inevitable economic expenses and losses either through immediate security investments in AI governance and controls, or through delayed implementation leading to competitive disadvantage and increased vulnerability to AI-enabled threats. Existing research demonstrates [36] how misaligned incentives lead to suboptimal security decisions, which is an important finding in the cybersecurity economics; our research extends this, arguing that even perfectly aligned incentives cannot prevent security compromises.

Moreover, research [37] on the economics of AI adoption highlighted how market pressures create "adoption races" where organizations must implement AI technologies to stay competitive. These races create "forced investment scenarios", namely, situations where both the investment and non-investment path will lead to cybersecurity-related economic losses. Empirical insights from a large-scale survey [38] support this notion, showing that these types of analyses do not consider the security implications of such races. With the proposed theoretical framework, we bridge this gap and explicitly model how competitive pressure creates cybersecurity dilemmas that cannot be resolved through traditional economic optimizations.

### C. Organizational Decision Theory Under Technological Pressure

Organizational decision theory in the context of cybersecurity has extensively studied how enterprises make strategic choices under uncertainty [39] [40] [41]. Cybersecurity leaders with decision making power, can settle for "good enough" choices, even if they are not perfect [42]. This is in alignment with the classical organizational behaviour theory, where decision makers operate under bounded rationality and incomplete information [42]. However, at the intersection of AI and cybersecurity, that compromise isn't always possible because none of the available options meet basic security needs [37]. Recent studies [43] [44] on organizational decision-making show that in many cases, decision-makers aim to make "satisficing" choices, namely, choices that are acceptable but not necessarily optimal. Nonetheless, AI security decisions can place organizations in situations where none of the options can even satisfy basic security standards. For example, when deploying LLMs within enterprise IT environments, security leaders face decisions where every option violates established security principles, thus, allowing broad LLM access creates data leakage risks, restrictive access renders the technology practically ineffective, and delayed deployment exposes the organization to competitive risks.

This phenomenon poses a challenge to the classic organizational decision model [45], which assumes that there is at least one minimally acceptable alternative. Our research extends this theory and examines scenarios where technological pressure eliminates the possibility of such satisficing choices. When it comes to technology adoption, organizations can delay decisions until a better option emerges. However, when it comes to AI security decisions extend beyond "forced choice" [46] scenarios (where organizations must select from undesirable alternatives) into zugzwang positions. Zugzwang captures both the forced nature of the choice and the inevitable deterioration of security position over time, creating increasing pressure for action despite inadequate alternatives.

### D. The AI Security Zugzwang Framework

The intersection of these three theoretical domains serves as the foundation for understanding the AI security zugzwang positions. It reveals how traditional assumptions about security decision-making demonstrate limitations when:

- Competitive pressure removes the option of non-adoption, whereas workforce productivity gaps widen between adopters and non-adopters
- Security implications of adoption cannot be fully understood or mitigated, yet productivity benefits accelerate the deployment
- Delay in decision-making itself creates security vulnerabilities, and simultaneously impacting workforce efficiency and talent retention [47].

Grounded on the theoretical synthesis of game theory, security economics and organizational decision theory, we propose a novel framework to understand the AI security zugzwang. It is characterized through the following three properties, which arise from our theoretical foundation analysis:

**1. Forced movement:** AI security zugzwang situations where maintaining the fully secure status quo is not viable, as opposed to traditional security decisions where organizations can maintain current positions. This property derives from both external pressures such as market competition, regulatory requirements and internal pressures such as technological obsolescence security debt and workforce productivity demands [25].

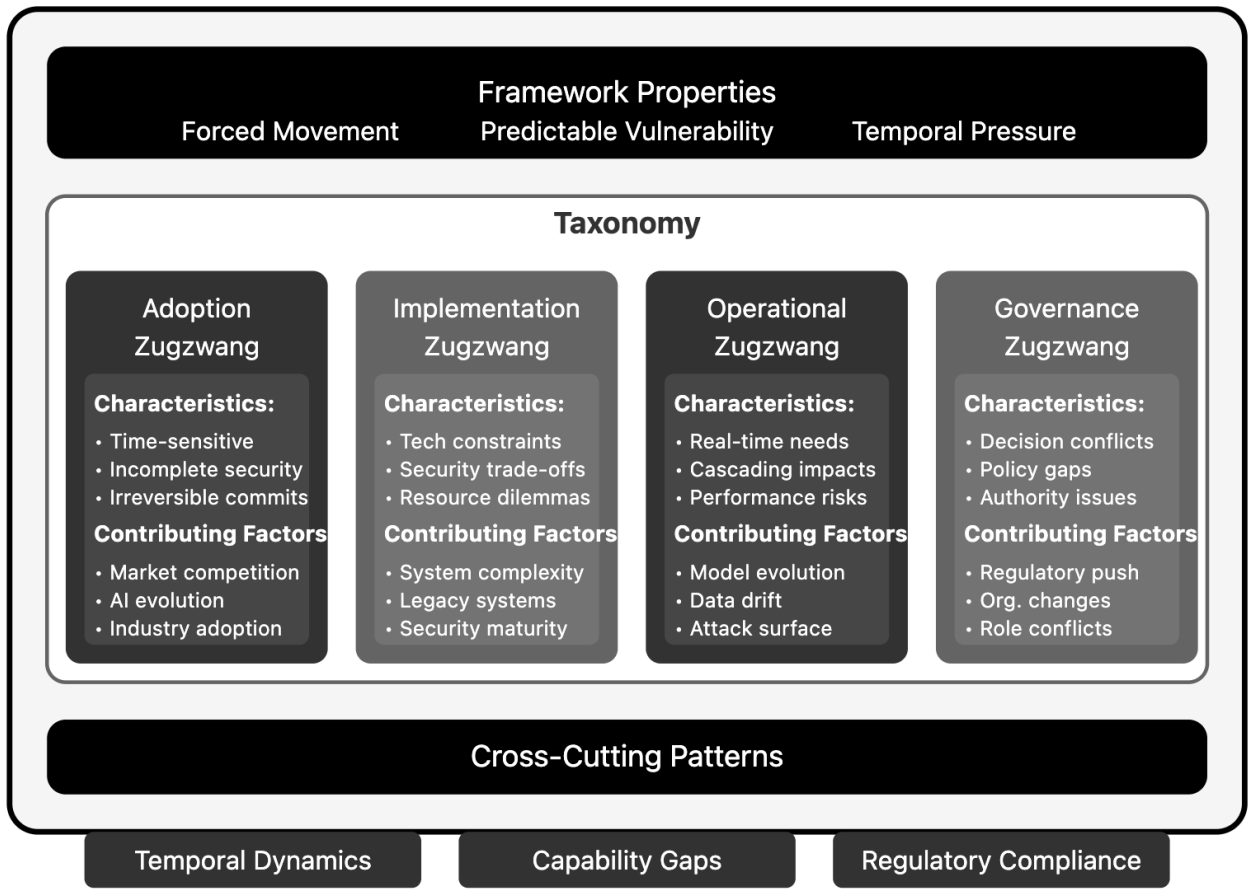

*Figure 1 - The AI Cybersecurity Zugzwang Framework and Taxonomy.*

**2. Predictable vulnerability creation:** Each possible move in an AI security zugzwang position creates new, identifiable security vulnerabilities. This is different than the traditional security trade-offs where positive outcomes are possible. A recent work on AI system confidentiality [48] demonstrated how even optimal security configurations introduce new attack surfaces.

**3. Temporal Pressure:** The security impact of decisions in AI security zugzwang positions worsens with delay, thus, creating urgency that further constrains decision-making. This is in alignment with work from scholars [49] on time-dependent security degradation but adds the dimension of competitive pressure unique to AI adoption.

These properties become apparent in various organizational contexts, leading to clear and distinct patterns of zugzwang positions that we categorize in our taxonomy. Most importantly, these patterns emerge regardless of organizational size, sector, or security maturity, which suggests that they are inherent to the current state of AI technology rather than implementation specific [50].

## IV. TAXONOMY OF AI SECURITY ZUGZWANG POSITIONS

To help organizations move from theory to practical application and operationalize our framework, we define a taxonomy of AI security zugzwang positions based on four primary categories and three cross cutting patterns. These positions represent distinct patterns of forced security moves that organizations face in AI adoption and implementation. The framework together with the taxonomy are visualised in Figure 1.

**1. Adoption Zugzwang**

In adoption scenarios, time pressure becomes the dominant force, creating situations where organizations must make security decisions with incomplete understanding of their implications. This pressure is significant in competitive industries where AI adoption rates directly impact market position. Therefore, organizations face this position when market pressures force AI adoption decisions, that inevitably create security vulnerabilities. The key characteristic here, is the impossibility of maintaining a secure status quo. For instance, when competitors adopt large language models (LLMs) for customer service, then organizations must either:

- Adopt LLMs and accept new attack surfaces related to prompt injection and data exposure [51], or,

- Maintain traditional systems and risk competitive disadvantage while facing increased attack sophistication from AI-enabled threats [52]

To identify adoption zugzwang positions, we develop Table 1 that details the characteristics and contributing factors. They oftentimes appear simultaneously, creating complex decision scenarios where security leaders and teams must figure out the best tactic to deployment pressure against security readiness.

*Table 1 - Adoption Zugzwang detailed characteristics and contributing factors.*

| Characteristics | Contributing Factors |
|---|---|
| Business units demanding AI deployment within 3-6 months | Competitors actively deploying similar AI capabilities in production |
| Critical security assessments pending while facing deployment deadlines | Weekly/monthly releases of new AI capabilities requiring security evaluation |
| Infrastructure changes requiring 6 or more months of security rework after deployment | Industry regulations requiring AI adoption for specific functions (e.g., fraud detection) |

### 2. Implementation Zugzwang

Implementation zugzwang positions emerge from the fundamental tension between security controls and system functionality. When organizations implement AI systems, they frequently face technical constraints that create impossible choices, for instance, implementing necessary security controls inevitably degrades critical AI system performance, or, maintaining performance would require to accept substantial security risks [53]. These tensions or even frictions sometimes, is enhanced by the relative immaturity of AI security tooling and the complexity of integrating these systems with existing security infrastructure. Security leaders and teams can use Table 2 to recognize implementation zugzwang scenarios through their characteristics and factors. These highlight how technical constraints create zugzwang positions, even when organizational intent and resources for secure implementation exists.

*Table 2 - Implementation Zugzwang detailed characteristics and contributing factors.*

| Characteristics | Contributing Factors |
|---|---|
| Security controls causing model accuracy to drop below acceptable thresholds | AI models requiring real-time processing across multiple data sources |
| Essential AI features blocked by existing security controls | Legacy security tools unable to monitor AI-specific threats (e.g., prompt injection) |
| Security teams lacking AI expertise while facing immediate protection requirements | Available security tools requiring 6 or more months of integration work for AI coverage |

### 3. Operational Zugzwang

These positions represent perhaps the most persistent category, which is characterized by the continuous tension between security requirements and operational demands. Organizations must constantly stabilize the need for model updates against supply chain security risks and manage access controls that either compromise security through excessive permission or degrade functionality through excessive restriction [54]. Another operational zugzwang example is when any decision to enhance visibility for security purposes, such as monitoring user prompts, leads to a potential compromise in user privacy, although opting not to increase visibility exposes the system to either AI-driven threats or data leakage [55]. Table 3 present the key characteristics and contributing factors of operational zugzwang positions. The persistence of these characteristics in daily operations indicates how AI security zugzwang becomes embedded in routine security management, thus, needs to continuously balance of competing security and operational demands.

*Table 3 - Operational Zugzwang detailed characteristics and contributing factors.*

| Characteristics | Contributing Factors |
|---|---|
| Security patches requiring immediate deployment without complete testing | AI models requiring weekly updates to maintain accuracy while lacking security validation processes |
| Security changes in one AI component breaking functionality in connected systems | Training data requiring constant updates while introducing potential poisoning vectors |
| Security monitoring causing unacceptable latency in critical AI operations | Each new AI feature expands the attack surface through new data inputs and API endpoints |

### 4. Governance Zugzwang

Governance zugzwang positions arise when organizations must make immediate decisions about AI security governance whilst lacking established frameworks. These positions can be better noticed when security governance structures cannot accommodate AI system requirements, however, operations cannot pause for governance development. For instance, when AI models need real-time security decisions, but existing approval processes create unacceptable delays, or when security teams must implement AI-specific controls although decision rights remain disputed between AI, IT, security teams, and risk committees. This category focuses more on the decision rights and organizational structures rather than system operations. Namely, when specific governance-related challenges emerge, such as, security teams lacking clear authority to respond to AI-specific threats, boards must set risk tolerances without complete understanding of security implications, and regulatory requirements demanding governance structures without providing implementation guidance. Table 4 outlines the characteristics and contributing factors that help leaders identify governance zugzwang positions. The presence of these characteristics shows governance challenges that need careful consideration of whether existing security governance approaches need to be adapted or supplemented with new frameworks for decision-making and accountability.

*Table 4 - Governance Zugzwang detailed characteristics and contributing factors.*

| Characteristics | Contributing Factors |
| --- | --- |
| Security decisions requiring immediate action while lacking established approval framework | Board/C-level-mandated AI initiatives conflicting with security assessments timelines |
| Critical security policies needing implementation before governance structures mature | Regulatory requirements demanding immediate governance changes while lacking implementation guidance |
| Security roles and responsibilities need immediate definition while facing organizational flux | Overlapping responsibility domains between AI teams, security teams, and risk committees creating decision paralysis |

5. **Cross-Cutting Patterns**

The analysis of AI security zugzwang positions shows three patterns that are applicable throughout all categories.

**Temporal dynamics,** when organizations delay decisions, their security position deteriorates at an accelerating rate, thus, creating what can be characterized as "security debt." The deterioration appears in concrete ways such as new AI model versions introduce vulnerabilities that compound over time, or security patches become obsolete within weeks, and delayed security decisions create cascading risks across interconnected systems. This deterioration is different from traditional technical debt in the sense that it cannot be addressed through incremental improvements or additional resources. Eventually, the "security debt" accumulation creates pressure for sub-optimal decisions [56].

**Capability gaps** is the second cross-cutting pattern. Organizations struggle to develop or acquire the security capabilities necessary to fully address AI-specific threats. These gaps are mostly found in areas of AI system monitoring, where traditional security tools prove inadequate for detecting model-specific anomalies, verifying AI system behaviour, or enhance visibility during user prompts [57]. Organizations also face significant challenges in finding and retaining security professionals with the expertise needed to understand both AI systems and security implications [58]. This expertise gap is important because it requires individuals who can "bridge" security practices with emerging AI-specific threats, whilst also having a basic understanding of the underlying machine learning models and their unique vulnerabilities. The challenge becomes even more complicated because of the difficulty of warranting supply chain security in AI systems, where vulnerabilities might exist in pre-trained models or third-party components [59].

**Regulatory compliance** is the third cross-cutting pattern that adds complexity throughout all zugzwang categories. Organizations being caught between conflicting regulatory requirements, especially when different jurisdictions impose incompatible constraints on AI system operation and data handling. This compliance issue becomes worse because of the fast-paced development of AI tools and capabilities, which is already outpacing the development of regulatory frameworks [60]. Ultimately, this results in a situation where compliance with one requirement can force violations of another, and thereby create a regulatory zugzwang that compounds technical and operational challenges.

## V. STRATEGIC MITIGATION AND POSITION MANAGEMENT

The analysis of zugzwang properties and patterns helps us to understand **"the why"** traditional decision frameworks do not account for such scenarios. An assumption within existing security decision frameworks is that organizations can choose between clear security alternatives or defer decisions until more information becomes available [61]. However, AI security zugzwang positions, challenge this assumption because of the:

- **Decision delay** accelerates the security risks, thereby creating a form of forced urgency. Contrary to traditional security decisions were waiting can provide more information or better options, AI security zugzwang positions worsen with delay. This deterioration occurs through competitive advantage loss, technical and security debt accumulation, and expanding attack surfaces [56].
- **Cascading effects of security decisions** that are impossible to fully predict or contain beforehand, which is caused by the interconnected nature of AI systems. When organizations implement security controls for AI systems, these controls inevitably affect multiple system aspects, and therefore create new vulnerabilities while addressing existing ones [55].
- **Technical evolution** of AI capabilities means that security decisions must be made against a backdrop of continuous technical change. This in turn poses a problem to the standard planning cycles, where security decisions are planned with a high degree of certainty, ultimately making it increasingly difficult to establish stable security positions [62].

To enhance AI security decision making and help on **"the how"** to manage zugzwang positions, our research identifies three practical requirements:

1. Organizations need systematic capabilities like cyber foresight [63] to spot developing zugzwang positions early. This means, conducting regular "security radar reviews" when, for example, multiple business units demand LLM deployments while basic security monitoring and prompt visibility are not in place. Moreover, isolated security metrics alone aren't enough; organizations need to combine metrics on business pressure, technical dependencies, and security gaps into unified probabilistic charts [64] that will enable them with the necessary helicopter view to translate the charts and achieve **early recognition**.
2. Oftentimes security architecture assumes more controls equal better security. However, zugzwang positions need "decision-space architecture", thus, security frameworks that plan for inevitable forced moves. This means designing modular security controls and predetermined security boundaries that can adapt without creating new vulnerabilities during AI adoption phases, therefore, achieving **architectural adaptation**.
3. Security teams can't manage zugzwang positions in silos. Organizations need interoperable cyber value chains that enable rapid security adjustments based on real-time risk and threat assessments [65]. This means tight integration between security teams performing threat analysis, business leaders making AI adoption decisions, and technical teams implementing controls. Although this approach might be resource-intensive in the beginning, it enables organizations to adapt security posture as zugzwang positions evolve, thereby achieve **cross-functional integration**. This interoperable way of working can be vastly improved onwards with the use of automation and become more efficient and effective.

To operationalize these requirements, we develop practical response tactics. Because of the rapid evolution and adoption of AI technology, these positions will become more frequent and complex. It is therefore imperative for organizations to build institutional expertise to recognize and respond to zugzwang scenarios past technical solutions. This expertise must extend beyond the security team to include business leaders who make AI adoption decisions and technical teams responsible for AI implementation. Therefore, clear articulation and mutual understanding of objectives between these teams becomes equally imperative.

### A. Decision Tactics and Their Implications

Our framework and taxonomy reveal three distinct tactical approaches to AI security zugzwang decisions, each addressing the previously identified limitations differently.

**The Minimization Tactic**

This tactic helps organizations to control their exposure by limiting AI system scope and functionality as a starting point. This approach reduces security risks initially, but it also creates "security illusions". Namely, situations where perceived risk reduction covers increasing vulnerability. For example, if organizations severely restrict their AI models access to data this will lead the models in becoming less effective at detecting actual security threats, thus, creating a different kind of security exposure. This tactic is most effective when organizations clearly define success criteria for their use cases upfront, thereby minimization tactic becomes the equivalent of the need-to-know principle.

**The Acceleration Tactic**

This tactic is used to respond to zugzwang positions with an accelerated AI security decision, as an attempt to "move through" a vulnerable position quickly. This tactic helps organizations in rapid need of deployment of security controls and aggressive AI adoption timelines. Although this approach helps organizations to maintain a competitive position, it often leads to security technical debt, namely, accumulated security vulnerabilities that become progressively difficult to address over time.

**The Adaptive Tactic**

The most sophisticated response tactic would be for organizations to develop adaptive security frameworks that can expand with changing conditions. This approach acknowledges the inevitability of zugzwang positions and focuses on building cyber resilience rather than prevention. However, it requires significant organizational maturity and resource commitment, therefore some organizations may find it impractical to begin with, but as the maturity and commitment increases initial steps towards adaptive cyber value chains can be achieved.

### B. Tactic Effectiveness and Organizational Impact

The effectiveness of these decision tactics varies based on the following dimensions arising from our literature review.

**Temporal considerations**

Minimization tactic provides immediate risk control, but the literature shows it becomes increasingly unsustainable. A recent survey on AI security adoption [66] demonstrated how delayed adoption leads to more severe security challenges when organizations eventually must implement AI capabilities. On the other hand, the acceleration tactic provides quick wins but accumulates technical debt, and adaptive approaches require longer implementation timeframes but offer sustainable security positions.

**Resource implications**

Each tactic needs different organizational investments. Minimization is the least resource-intensive initially but often requires crisis-driven investments later. Acceleration needs rapid resource mobilization for implementation and incident response. The adaptive tactic, while theoretically optimal, requires sustained investment in expertise, governance frameworks, and risk management capabilities. Many organizations will find this tactic prohibitive, thereby will be forced into less optimal tactics despite understanding the risks.

**Organizational learning**

All tactics create different organizational learning trajectories. Organizations following the acceleration tactic must develop rapid response capabilities, but this comes with lack of deep understanding of their security positions. Minimization delays critical capability building and adaptive tactic builds sophisticated security capabilities but requires longer learning curves and at the cost of increased initial vulnerability.

Ultimately, each tactic addresses the framework properties and patterns differently, thus, creating distinct implications for security posture and business impact, as shown in Table 5.

*Table 5 - Comparative analysis of AI Security Zugzwang tactics and their organizational implications.*

|  |  | **Minimization Tactic** | **Acceleration Tactic** | **Adaptive Tactic** |
|---|---|---|---|---|
| **Security Impact** | | • Initial control due to restricted AI access and data boundaries<br>• Shadow AI creates unmonitored security exposures<br>• Limited visibility into AI-specific threats<br>• Reactive incident response capabilities<br>• Increasing blind spots as AI adoption grows | • Rapid security control deployment<br>• Early detection of AI attack patterns<br>• Production-discovered vulnerabilities<br>• Compressed security testing cycles<br>• Real-time security adaptation needs | • Systematic security implementation<br>• Structured vulnerability management<br>• AI monitoring across the board<br>• Proactive threat detection<br>• Evolving security controls |
| **Business Impact** | | • Significant productivity differential compared to AI-enabled competitors [cite Harvard study]<br>• Technical talent retention challenges<br>• Manual process overhead<br>• Loss of AI-driven market opportunities<br>• Growing technical debt from delayed adoption | • Early productivity gains from AI adoption<br>• Market positioning advantage<br>• Higher initial security investment<br>• Increased incident response load<br>• Technical debt accumulation | • Balanced productivity growth<br>• Sustainable – leading market position<br>• High ongoing investment needs<br>• Effective talent retention<br>• Coordinated adoption process |
| **Resource Requirements** | *Initial* | Low | High | High |
| | *Ongoing* | High due to crisis-driven catch-up | Moderate-High | High |
| | *Security Monitoring* | Basic | Advanced | Comprehensive |
| | *Governance* | Traditional frameworks | Rapid development | Mature frameworks |
| | *Expertise* | Standard security skills | AI security specialists | Deep AI security expertise |
| **Risk Profile** | *Short-term* | Controlled | High | Moderate |
| | *Medium-term* | Growing exposure | Moderate | Managed |
| | *Long-term* | Critical | Manageable | Low |
| | *Primary risks* | • Shadow AI adoption<br>• Competitive disadvantage<br>• Security capability gaps | • Security gaps<br>• Control effectiveness<br>• Resource strain | • Resource intensity<br>• Implementation complexity<br>• Coordination overhead |
| **Success Requirements** | | • Clear use case scope boundaries<br>• Strong policy enforcement & stakeholder alignment/synergy<br>• Regular risk assessments<br>• Shadow AI monitoring | • Robust incident response<br>• Rapid patching capability<br>• Technical debt tracking<br>• Resource availability | • Mature security processes<br>• Strong governance across IT & security<br>• Cross-team coordination & collaboration<br>• AI security expertise |

## C. Decision Flowchart and Real-World Application

To operationalize the framework, taxonomy and tactics, we draw a decision flowchart (Figure 2) that guides organizations through zugzwang position identification, analysis and response selection. The flowchart is based on the framework properties, taxonomy categories and tactics, and provides a structured decision path for security leaders.

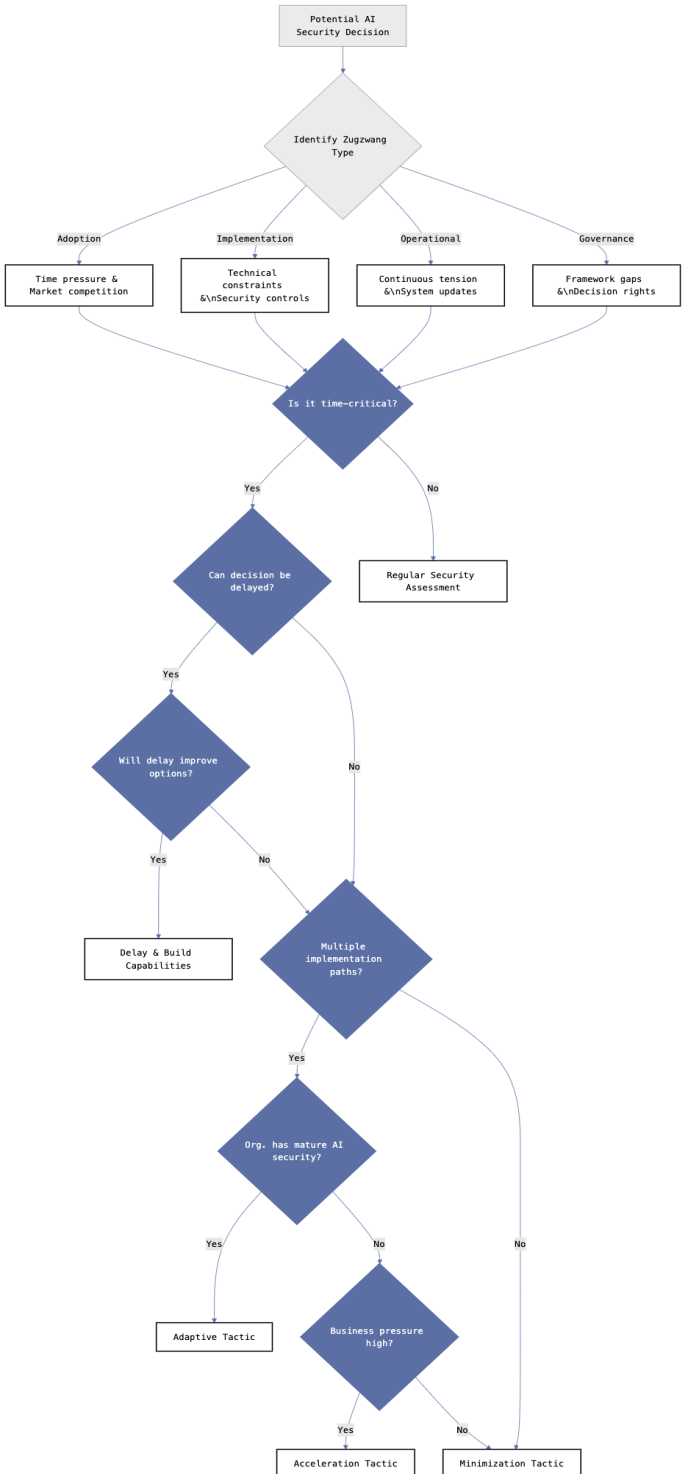

*Figure 2 - Decision Flowchart.*

We examine a widespread zugzwang position, namely, Microsoft Copilot adoption. This case demonstrates the framework's practical application and value when guiding strategic response. The zugzwang position emerged when business units demanded rapid deployment to enhance developer productivity and streamline operations through AI-powered features, despite security teams identifying significant risks due to scattered sensitive data throughout SharePoint sites, or team's channels, and the absence of proper security controls. In the anonymized case, internal metrics showed 40% of developers were using unauthorized AI coding assistants, creating immediate security exposure through potential data leakage.

Analysis through our taxonomy identifies this as primarily an adoption zugzwang, driven by intense market competition and time pressure from business demands. The time-critical nature becomes evident through the immediate productivity impact and risk of shadow IT adoption if formal deployment is delayed. When evaluating delay possibilities, the framework reveals that standard security approaches of postponing implementation until full security controls are in place would be ineffective, given the business's fast-tracking of the initiative and the scattered nature of existing data across collaboration platforms.

The framework then guides the security leaders through implementation options, highlighting multiple possible paths starting with phased rollouts, department-based deployment, and feature-limited implementation. However, the organizational AI security maturity assessment reveals significant gaps. For instance, lack of AI prompt monitoring capabilities through the SOC, which leads into insufficient SOC readiness. Additionally, the absence of AI-specific security controls leads into ineffective and inefficient incident management. Combined with the high business pressure evidenced by executive mandate and developer productivity demands, the framework then directs towards the acceleration tactic.

This tactical direction is valuable as it acknowledges the reality of the situation and provides a structured approach to managing risk. Instead of either blocking deployment (which would likely fail) or allowing uncontrolled implementation, the framework guides security leaders towards a rational response, namely, rapid deployment with parallel security capability development. This approach aligns with business needs and maintains security oversight and progressive control implementation.

Thus, the framework serves as a practical tool for security leaders, helping them steer complex AI security decision making with confidence, clear and documented decision points, and tactical options. It validates the zugzwang nature of the situation -where neither full security nor complete blocking is viable- and provides a structured path forward that balances innovation and risk management.

## VI. DISCUSSION AND IMPLICATIONS

A fundamental question emerges when examining the AI security zugzwang. Namly, how does this differ from traditional security decision-making? Organizations normally accept and manage security risks, but our research shows differences in AI security zugzwang positions.

First, the traditional security theory assumes organizations can maintain existing security positions and evaluating alternatives. Our framework shows how AI security zugzwang invalidates this assumption through forced moves and deteriorating positions. AI decisions create exponentially compounding security issues that exceed conventional security controls, as opposed to the standard scenarios where risk acceptance follows linear patterns.

Second, the temporal dynamics of zugzwang positions challenge existing security risk assessment approaches. We found that traditional models assume better information leads to better decisions, and our analysis shows how risk acceleration in AI security outpaces organizations' ability to gather and process security-relevant information. This temporal pressure, evident in our Copilot case study, requires a new approach to security assessment under the pressure of time, much like in chess.

Third, conventional risk management treats risk as the product of likelihood and impact and most of the times this is visualized and communicated through risk matrices, which is insufficient for zugzwang positions [67]. That is because of these positions create inevitable negative outcomes rather than binary events like "will this happen or not" events, and their interconnected nature defies the traditional compartmentalized risk assessment [68]. The adaptive tactic specifically highlights how traditional risk frameworks do not to capture the complexity of AI security decisions.

Our research suggests that probabilistic methods coupled with quantitative modelling are more suitable to capture and manage zugzwang positions. Monte Carlo simulations, for instance, can better capture the cascading effects and complex interactions we identified in our taxonomy. Such methods would allow organizations to model multiple interaction paths, incorporate temporal dynamics, and account for uncertainty mathematically. However, the novelty of AI security zugzwang positions presents a significant challenge, namely, the lack of historical data for probability estimation.

These findings suggest four core implications for organizations:
1. Security assessments must evolve beyond the outdated traditional impact/likelihood risk matrices to quantitative methods that account for zugzwang positions
2. Risk models must become dynamic rather than static and account for changing conditions
3. Probabilistic approaches are essential to understand the cascading effects identified in our taxonomy
4. Uncertainty must be explicitly incorporated into risk calculations, especially when applying different tactical responses.

## VII. FUTURE RESEARCH DIRECTION

The identification of AI security zugzwang as a distinct phenomenon opens several future research directions. These directions derive from the limitations of our current understanding, as well as the rapidly evolving nature of AI technology and its security implications. An immediate direction highlighted by this research is the development of quantitative models that can better capture zugzwang positions. Our framework and taxonomy provide a qualitative understanding, however, the development of mathematical models that can predict the evolution of these positions would significantly enhance the decision-making capabilities of cybersecurity leaders. Moreover, the development of specific metrics for measuring the "depth" of zugzwang positions would be another interesting research direction. Namely, understanding how to quantify the severity of these positions, and their likely impact on organizational security, would provide valuable decision support tools. Lastly, research on how AI security zugzwang positions propagate through technological ecosystems and supply chains could yield valuable insights, since the interconnected nature of AI systems suggests that zugzwang positions create ripple effects across organizational boundaries.

## VIII. CONCLUSIONS

The concept of AI Security Zugzwang provides a novel framework to understand and manage the unique security challenges organizations face in AI adoption and implementation. We demonstrated that these positions fundamentally differ from traditional security trade-offs, as they represent scenarios where maintaining current security postures becomes impossible, and every move inevitably creates new vulnerabilities. Ultimately, we make three core contributions: (1) a theoretical framework characterizing AI security zugzwang through forced movement, predictable vulnerability creation, and temporal pressure properties; (2) a practical taxonomy to identify and categorize zugzwang positions across adoption, implementation, operational, and governance contexts; and (3) tactics to manage these positions, which we validated through real-world application. Our findings suggest that effective management of AI security zugzwang requires organizations to move from traditional security decision-making models towards probabilistic approaches that account for temporal dynamics and cascading effects. Future research should focus on developing quantitative methods to model and predict the evolution of zugzwang positions mathematically and precisely.